Ultra-low-resistivity nitrogen-doped p-type Cu$_2$O thin films fabricated by reactive HiPIMS

Jiří Rezek*, Jan Koloros, Jiří Houška, Radomír Čerstvý, Stanislav Haviar, Jemal Yimer Damte, David Kolenatý, Pavel Baroch

Department of Physics and NTIS, European Centre of Excellence, University of West Bohemia in Pilsen, Univerzitní 8, 301 00 Pilsen, Czech Republic

*Corresponding author, Tel.: +420 377632269, E-mail address: jrezek@ntis.zcu.cz

Abstract

We have successfully fabricated the nitrogen-doped cuprous oxide thin films on the amorphous standard soda-lime glass by reactive high-power impulse magnetron sputtering. The energy of film-forming particles was controlled by the value of pulse-averaged target power density, which has a significant impact on the elemental composition, structure and optoelectrical properties of the films. We have shown that the high-energy regime is more suitable for preserving Cu$_2$O structure and leads to continuous substitution of oxygen by nitrogen compared with the low-energy regime. Moreover, in the high-energy regime, it is possible, to some extent, to independently control the electrical resistivity and optical properties. The electrical resistivity decreases down to ≈ 5 ×10$^{-2}$ Ωcm at the optical band gap 2.0-2.3 eV. Special attention is paid to the formation of nitrogen molecules and their ability to form shallow acceptor states. Experimental results supported by our DFT calculations indicate that N$_2$ replacing Cu in the Cu$_2$O lattice is one possible (but not the only possible) acceptor. We have also found that the formation of nitrogen molecules is preferred in a high-energy regime.

1. Introduction

One of the challenging scientific topics of today is finding a suitable p-type TCO that would at least approach the optoelectronic properties of the n-type counterpart [1]. Finding such p-type material is a necessary condition for the further sustainable growth of civilization. Realizing p-n junctions using transparent conductive materials will enable the development of a new generation of invisible electronics, contribute to reducing the energy requirements of various optoelectronic devices or lead to the production of more efficient solar cells.

Transparent conductive materials based on $Cu_2O$ appear to be among the most promising. This is mainly due to the abundance of elements used, their non-toxicity and interesting optoelectronic properties. One of the many applications of $Cu_2O$ is as an absorption layer for solar cells. It is because of a proper optical band gap of 2.0-2.6 eV [2–4] and the possibility of reaching relatively high hole mobilities [5–7]. It turns out that $Cu_2O$-based solar cells have one of the lowest cost-per-kWh ratios of any known material [8]. However, as with most other p-type TCOs, the main problem with $Cu_2O$ is its relatively high resistivity. Several strategies have been used and investigated to decrease this. Namely, deposition at elevated temperatures [9], post-deposition annealing [10,11], or post-deposition laser treatment [3], which have been proven to increase the mobility of holes. Even though the concentration of holes mostly decreases at the same time (healing of defects in the layer that serve as acceptor states), the overall conductivity of the layer increases. Another strategy is to dope $Cu_2O$ with other elements to create additional acceptor states near the valence band maximum (VBM). Successful doping of $Cu_2O$ with elements such as B, Li or Na was reported [12–14], which led to a significant increase in hole concentration and electrical conductivity. Special attention is then paid to doping $Cu_2O$ with nitrogen, an abundant and non-toxic element. The role of nitrogen in the $Cu_2O$ layer was described, e.g. by Lai et al., where nitrogen was added during rf-magnetron sputter deposition, resulting in a decrease in the electrical resistivity down to

≈ $1\times10^{-1}$ Ωcm for optimal nitrogen partial pressure [15]. Other authors also reported similar results [16–18]. It was generally believed that the increase in hole concentration observed in all the above-mentioned works is caused by a nitrogen molecule substituting a copper atom in the $Cu_2O$ lattice ([2,4]). However, in a recent in-depth study of Soltanmohammad and Nilius, it seems that the main mechanism is rather the substitutional replacement of oxygen by atomic nitrogen in the $Cu_2O$ matrix [19].

Although reactive high-power impulse magnetron sputtering (r-HiPIMS) is an advanced magnetron deposition method and has been successfully applied to the deposition of various TCOs [20,21] and even to the deposition of $Cu_2O$ [22], the potential of this method to prepare nitrogen-doped $Cu_2O$ has not been explored. The advantage of this technique is the possibility of influencing, to some extent, the energy and composition of the particle flux forming the growing film. This was demonstrated, for example, during HiPIMS deposition of IGZO [23] or AZO [24] layers, when changing the elemental composition and optoelectronic properties, was possible by adjusting the pulse target power density. In this publication, we have demonstrated that r-HiPIMS is a very suitable method for effective nitrogen doping of $Cu_2O$ films and, to some extent, for independent tuning of electrical/optical properties. The prepared p-type $Cu_2O$:N films exhibited an extraordinarily low resistivity below $5\times10^{-2}$ Ωcm while maintaining p-type character. Moreover, we found the role of the nitrogen molecule as one of the possible acceptors is probably still not ruled out.

## 2. Experimental setup

### 2.1. Film preparation

Cu-O-N thin films were deposited using reactive r-HiPIMS with a copper target (99.99% purity), with a diameter of 100 mm and a thickness of 6 mm (area, $A_t$, of 78.54 cm$^2$), in an argon–oxygen–nitrogen gas mixture. A scroll and turbomolecular pump were used to evacuate the chamber to a base pressure of approximately $6 \times 10^{-4}$ Pa. The schematic picture of our experimental apparatus with a more detailed description could be found in our recent work [3]. The working pressure during the deposition was controlled in two stages. In the first stage, a constant pressure of 0.5 Pa (Ar + N$_2$) was maintained by adjusting the pumping speed. Keeping the argon and nitrogen flow constant, the required nitrogen fraction in Ar + N$_2$ mass flow, $f_{N2}$, was set. In the second stage, the oxygen was added, and the oxygen partial pressure, $p_{ox}$, varied from 100 mPa to 340 mPa. The magnetron was driven with a pulsed power supply (Melec GmbH) with the rectangular-shaped voltage pulse on-time $t_{on}$ was 100 μs. The value of the off-time ($t_{off}$) was set to obtain the required pulse-averaged target power density $S_{da} = \frac{P}{d_c \times A_t}$, where $P$ is the average power (500W for all cases) and $d_c = t_{on}/(t_{on}+t_{off})$ is the duty cycle. The soda-lime glass (38 × 26 × 1 mm$^3$) and Si (100) wafers (26 × 15 mm$^2$) were used as substrates and were cleaned in isopropyl alcohol and distilled water before deposition (10 min for each process). The target-to-substrate distance was 100 mm, and the substrates were heated to 190 °C. The deposition time was between 110 s and 150 s, resulting in a film thickness of 160-250 nm.

### 2.2. Film characterization

The elemental composition of the selected Cu-O-N films was measured by wave dispersive spectroscopy (WDS) carried out in a scanning electron microscope (Hitachi, SU-70). Measurements were carried out at an accelerating voltage of 7 kV. Cuprite (Cu$_2$O) was used as the standard for copper and oxygen, while ZrN (sputtered, 50:50 ratio) served as the standard for nitrogen. Calibration and quantification were conducted using the PROZA method, neglecting the influence of the underlying silicon substrate. The standards and the analyzed samples were coated with a 14 nm carbon layer to minimize charging effects and improve surface conductivity during analysis. To characterize the phase composition of thin films, we used a Raman spectroscopy (Horiba Jobin Yvon LABRAM HR Evolution Raman spectroscope), with a laser wavelength of 532 nm. The structure of the films was characterized by X-ray diffraction (XRD) using a PANalytical X'Pert PRO MPD diffractometer working in the Bragg-Brentano geometry using a CuKα (40 kV, 40 mA) radiation. The electrical resistivity, $\rho$, was measured by the standard four-probe method. The optical band gap, $E_g$, was estimated using the Tauc plot method, which involves analyzing the optical absorption data. The transmittance $T$ and reflectance $R$ of the thin films were measured. The absorption coefficient $\alpha$ was calculated as $\alpha = -\left[ln\frac{T}{1-R}\right]\frac{1}{t}$, where $t$ is the thickness of the thin film layer. For direct band gap materials, the relationship $(\alpha h\nu)^2 = (h\nu - E_g)$ was applied, where $h\nu$ is the photon energy. A plot of $(\alpha h\nu)^2$ versus $h\nu$ was used to determine $E_g$. Photoelectrochemical (PEC) measurements were conducted in a custom cell made from chemically inert polytetrafluoroethylene (PTFE), with internal dimensions of 20 × 17 × 30 mm³. Optical illumination and electrochemical contact were defined by a 6 mm circular window sealed by the sample. The working electrode consisted of a Cu-O-N layer on the Si substrate. A spiral platinum wire, positioned 15 mm from the working electrode, served as the counter electrode, while a standard Ag/AgCl reference electrode was placed ~ 6 mm away (corresponding to +0.20 V vs SHE and +0.53 V vs RHE). The supporting electrolyte was 0.1 M $Na_2SO_4$ (pH 5.66). Illumination was provided by a xenon arc lamp with an AM1.5G filter, calibrated to 100 mW·cm⁻² at the sample surface, and delivered through a quartz window. Electrochemical measurements were carried out using a Squidstat Plus potentiostat (Admiral Instruments, USA). Open-circuit potential (OCP) measurements under dark and illuminated conditions were used to evaluate the semiconductor properties. The refractive index ($n$) and extinction coefficient ($k$) have been measured by spectroscopic ellipsometry using the J. A. Woollam Co. Inc. VASE instrument. The measurements were performed in reflection using angles of incidence of 55°, 60° and 65° in the wavelength ($\lambda$) range from 300 to 2000 nm. The optical model included the glass substrate, Cu-O-N layer and a surface roughness layer. The imaginary part of the permittivity ($\varepsilon_2$) of Cu-O-N has been expressed by a sum of Cody-Lorentz formula (oscillator)

with Lorentz oscillators (see the detailed discussion in Results). The formula for the former reads

$$\varepsilon_2 = L(E)G(E) \text{ for } E \geq E_g + E_t \quad \text{(Eq. 1a)}$$

and

$$\varepsilon_2 = L(E_g + E_t)G(E_g + E_t)\frac{(E_g+E_t)}{E}exp(\frac{E-E_g-E_t}{E_u}) \quad \text{for } E < E_g + E_t \quad \text{(Eq. 1b)}$$

where

$$L(E) = \frac{AB^2 E_n E}{(E^2-E_n^2)^2+B^2E^2} \text{ (Lorentz oscillator in one of its possible forms)} \quad \text{(Eq. 2)}$$

and

$$G(E) = \frac{(E-E_g)^2}{(E-E_g)^2+E_p^2} \text{ (Cody form of the variable band edge function)} \quad \text{(Eq. 3)}$$

where $A$ is the amplitude, $B$ is the broadening (damping), $E_n$ is the energy which the Lorentz oscillator is centered at, $E_p$ is the transition energy between the onset behavior and the Lorentz oscillator behavior, $E_g$ is optical gap in the narrow sense of the word, $E_g+E_t$ is the optical gap in a broader sense of the word (including non-zero density of electronic states close to its edges) representing the transition energy between band-to-band transitions and the exponential Urbach absorption tail and $E_u$ represents the shape of this tail.

2.3 DFT calculations

DFT calculations were performed using VASP with the PBE-GGA functional (Perdew et al.) [25] and the PAW method (Kresse and Joubert) [26]. A 2 × 2 × 2 $Cu_2O$ supercell was used, with a 550 eV cutoff energy and a 6 × 6 × 6 Monkhorst-Pack k-point mesh. Structures were relaxed until forces were below 0.005 eV/Å and energy convergence was within 1e$^{-7}$ eV. DFT-D3 dispersion corrections were included (Hassani et al.)[27]. Nitrogen atoms and nitrogen molecules were doped at interstitial, Cu, and O sites. Formation energies were calculated as:

Interstitial: $E_f = E_{Cu_2O+N} - E_{Cu_2O} - E_{\frac{N_2}{2}}^{tot}$,

Substitutional N (example for Cu-site): $E_f = E_{Cu_2O+N} - E_{Cu_2O} - \frac{E_{N_2}^{tot}}{2} + \frac{E_{tot}(Cu\ bulk)}{N_{Cu}}$ and

Substitutional N$_2$ (example for Cu-site): $E_f = E_{Cu_2O+N_2} - E_{Cu_2O} - E_{N_2}^{tot} + \frac{E_{tot}(Cu\ bulk)}{N_{Cu}}$

Where $E_{Cu_2O+N}$ is the total energy of supercell with one nitrogen atom, $E_{Cu_2O}$ is the total energy of supercell without defects, $E_{N_2}^{tot}$ is the total energy of nitrogen molecule, $E_{tot}(Cu\ bulk)$ is the total energy of the Cu supercell and $N_{Cu}$ is the number of Cu atoms in that supercell.

3. Results and discussion

3.1. Elemental composition

The elemental composition of the Cu-O-N layers prepared in the high-energy regime (at an oxygen partial pressure $p_{ox}$ = 260 mPa) as a function of the nitrogen fraction in the Ar + N$_2$ mixture, $f_{N2}$, is shown in **Fig. 1a**. We can see that the copper content remains practically constant (approx. 65 at.%) independent of the $f_{N2}$ value. Furthermore, a gradual substitution of oxygen by nitrogen is evident. Here, the previously reported advantage of the high pulse-averaged target power density, $S_{da}$, is utilized [28]. During this regime, the high electron temperature in the plasma discharge effectively dissociates nitrogen molecules to atoms with a higher sticking coefficient and nitrogen is more easily incorporated into the layer. This is also clear from **Fig. 1b**: for $S_{da} \leq 700$ Wcm$^{-2}$ the nitrogen content in the films is around ≈ 2 - 3 at. %, when the increase of $S_{da}$ to 1000 and 1300 Wcm$^{-2}$ the nitrogen content increases to ≈ 4 and 6 at. %, respectively.

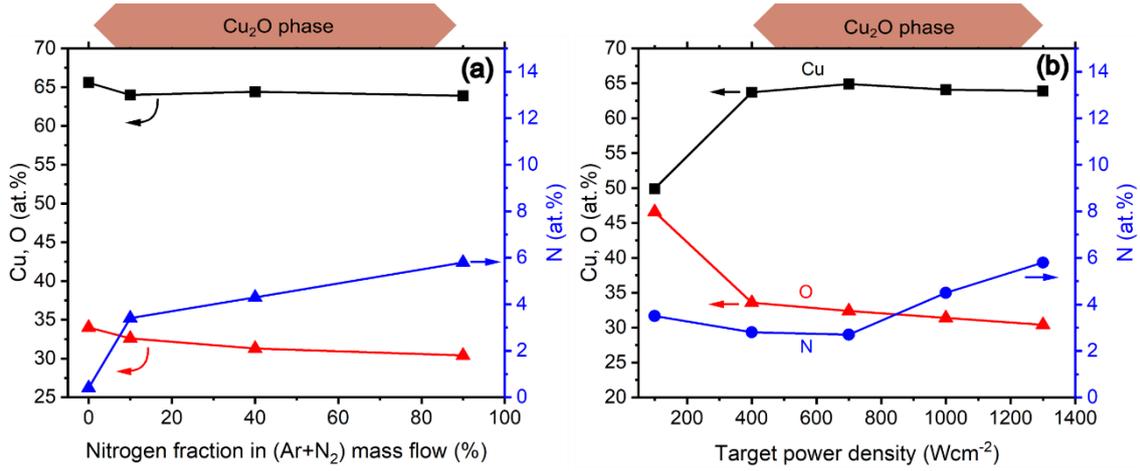

Fig. 1 (a) The elemental composition of Cu-O-N films fabricated at constant pulse-averaged target power density $S_{da}$ = 1300 Wcm$^{-2}$, $p_{ox}$ = 260 mPa and as a function of nitrogen fraction in Ar + N$_2$ mass flow, $f_{N2}$. (b) The elemental composition of Cu-O-N films fabricated at $p_{ox}$ = 260 mPa, $f_{N2}$ = 90% as a function of $S_{da}$.

To demonstrate the fundamental differences in the growth conditions of Cu-O-N layers in the low-energy ($S_{da}$ = 100 Wcm$^{-2}$) and high-energy ($S_{da}$ = 1300 Wcm$^{-2}$) regimes, a series of layers were prepared at different oxygen partial pressures, $p_{ox}$. To highlight the observed effects, the layers were prepared at the highest $f_{N2}$ = 90%. At first glance, a clear difference between the two regimes is visible. In the low-energy regime (**Fig. 2a**), there is a noticeable decrease in the copper content in the layer from about 63 at.% to 50 at.% with an increase in $p_{ox}$ from 160 mPa to 260 mPa. In contrast, in the high-energy regime (**Fig. 2b**), the copper content is practically constant in a very wide interval of 160 - 340 mPa. In the low-energy mode, there is also a sharper increase in the oxygen content in the layer (with increasing $p_{ox}$), which leads to an undesirable gradual transition to a material with a higher oxidation state of copper (Cu$_4$O$_3$, CuO or their combinations, see also below). Therefore, the high-energy mode is clearly more suitable for fine-tuning the nitrogen content in Cu$_2$O:N films.

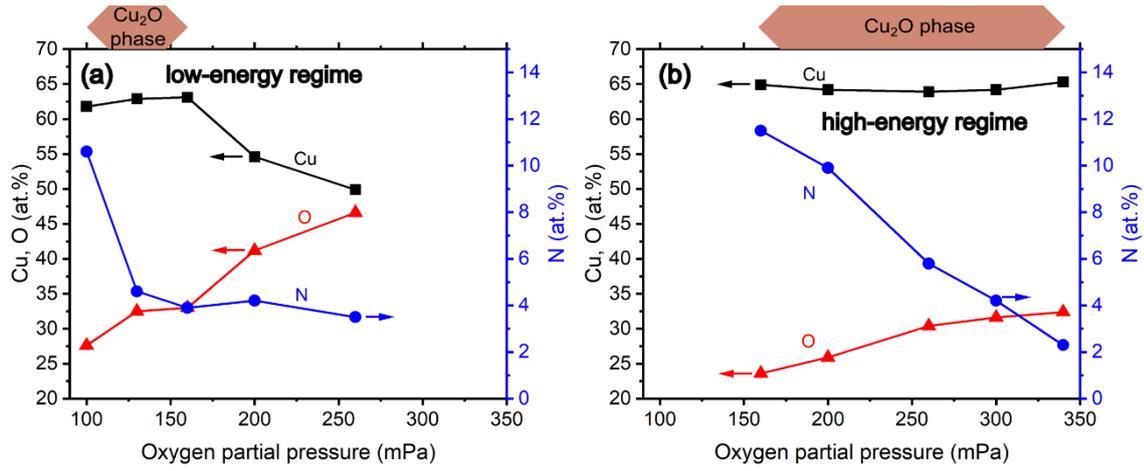

Fig. 2 (a) The elemental composition of Cu-O-N films fabricated in low-energy mode ($S_{da}$=100 Wcm$^{-2}$) at constant $f_{N2}$ = 90% as a function of oxygen partial pressure, $p_{ox}$. (b) The elemental composition of Cu-O-N films fabricated in high-energy regime ($S_{da}$ = 1300 Wcm$^{-2}$) at $f_{N2}$ = 90% as a function of $S_{da}$. The intervals when only Cu$_2$O phase is detected are marked.

### 3.2. Structural properties

Evolution of Raman spectra for the films prepared at different $f_{N2}$ being in the range of 0 - 90% (at constant $p_{ox}$= 260 mPa and $S_{da}$ = 1300 Wcm$^{-2}$) are depicted in **Fig. 3**. We can observe several Raman peaks. The peak at about 145 cm$^{-1}$ corresponds to the T$_{1u}$ cuprite mode, and oxygen vacancies activate it. The intensity of this peak decreases with increasing $f_{N2}$, indicating a decrease in the number of oxygen vacancies in the films [29]. Other significant peaks (around 208, 531 and 608 cm$^{-1}$) prove that only the Cu$_2$O phase is present in all films containing up to ≈ 6 at.% of nitrogen. XRD results also double-checked this, where only two prominent peaks corresponding to the cubic Cu$_2$O phase were detected for the films prepared at $f_{N2}$= 0 and 90% (see **Fig. 4a**). The positions of the most significant peaks, 531 and 608 cm$^{-1}$ for the film without nitrogen, are slightly shifted (10 - 15 cm$^{-1}$) to the lower wavenumber values when the nitrogen is introduced. Also, we can see the broadening of those two peaks with increasing nitrogen content, probably due to the higher density of crystal defects. Finally, there is a continuous increase of the peak located around 2250 cm$^{-1}$ for the films with $f_{N2}$ ≥ 10%. This peak corresponds to the nitrogen molecule [16].

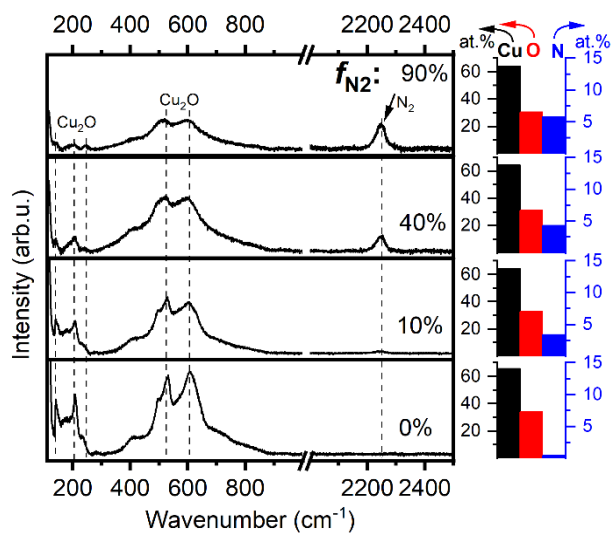

Fig. 3 A Raman spectrum of the Cu-O-N films fabricated in the high-energy regime at constant $p_{ox}$ =260 mPa as a function of $f_{N2}$. The elemental composition of individual films is plotted for better clarity.

Raman spectra of the films prepared at different values of $S_{da}$, i.e. at varying energies of film-forming particles are shown in **Fig. 5**. The film prepared at the lowest $S_{da}$ = 100 Wcm$^{-2}$ exhibits CuO structure (which corresponds to elemental composition of Cu$_{50}$O$_{47}$N$_3$), other films exhibit standard Raman patterns of Cu$_2$O discussed above. A slight broadening and shifting of the peaks to the lower values of wavenumbers is caused by an increasing nitrogen concentration. The intensity of the peak around 2250 cm$^{-1}$ (N$_2$ molecule) also increases with the $S_{da}$ value. Moreover, it is continuously shifting to the lower wavenumbers, probably caused by different bonding mechanisms of the nitrogen molecule under various growing conditions [16]. Interestingly, although films prepared at $S_{da}$ = 100-700 Wcm$^{-2}$ have practically the same nitrogen concentration ( ≈ 3 at.%) and the films prepared at $S_{da}$ = 400 and 700 Wcm$^{-2}$ even the very similar Cu and O content (≈ 64 at. % and 33 at. %, respectively), there is a visible increase in N$_2$ molecule Raman peak (not detected for the $S_{da}$ = 100 Wcm$^{-2}$). This indicates that the higher energy of film-forming particles favours the formation of N$_2$ molecules.

The next paragraph will discuss significant and qualitative differences in the structure of the films prepared at low- and high-energy conditions. **Fig. 6** shows Raman spectra of films prepared at both regimes and at different oxygen partial pressures. In the low-energy regime, the single-$Cu_2O$ structure is presented in the range of $p_{ox}$ from 100 to 160 mPa. Above that threshold, there is a mixture of $Cu_4O_3$+CuO phases ($p_{ox}$=200 mPa) or a single-CuO phase film ($p_{ox}$=260 mPa). For the high-energy regime, the situation is quite different. All films in the studied $p_{ox}$ interval (160 - 340 mPa) remain single-phase $Cu_2O$. This was also checked by XRD (**Fig. 4b**). From this point of view, the high-energy regime is more suitable for the preparation of $Cu_2O$ films without undesired admixture or other phases like $Cu_4O_3$ or CuO. The main reason is probably that, due to high-energy and high-density plasma, the target is prevented from being covered by copper (sub)oxides, and therefore, there is no significant drop in sputtering yield, as probably is the case in the low-energy regime. This means that there are still enough copper atoms ready to form $Cu_2O$ film, and the transition to $Cu_4O_3$ or CuO will occur only at very high oxygen partial pressures (beyond our experimental values). This behaviour was also observed in other work dealing with reactive sputtering at high target power densities [30]. Another interesting finding is the different trend in the $N_2$ molecule Raman peak intensity. While in the low-energy regime we observe a significant decrease in the intensity of the $N_2$ molecule signal with increasing $p_{ox}$ (i.e. with more oxygen in the chamber), practically down to zero for $p_{ox} \geq$ 200 mPa, in the high-energy regime, the opposite trend is observed. We see a gradual increase in the peak intensity from 160 mPa to 300 mPa. This is despite the fact that the total nitrogen concentration in the layer with $p_{ox}$ decreases (from approximately 10 at.% to 2 at.% with an increase in pox from 160 to 300 mPa). The mechanism of this behaviour is not yet completely clear, but it is certain that it cannot be explained only by the elemental composition, which would possibly favour the formation of the $N_2$ molecule. An example is the comparison of a layer formed in the low-energy regime at $p_{ox}$=160 mPa and a layer formed in the high-energy regime at $p_{ox}$=300 mPa: both layers show practically identical elemental composition ($Cu_{63}O_{33}N_4$ vs $Cu_{64}O_{32}N_4$), but while the peak of the $N_2$ molecule in the "low-energy" layer is almost negligible (**Fig. 6a**), in the "high-energy" layer it is very intense (**Fig. 6b**).

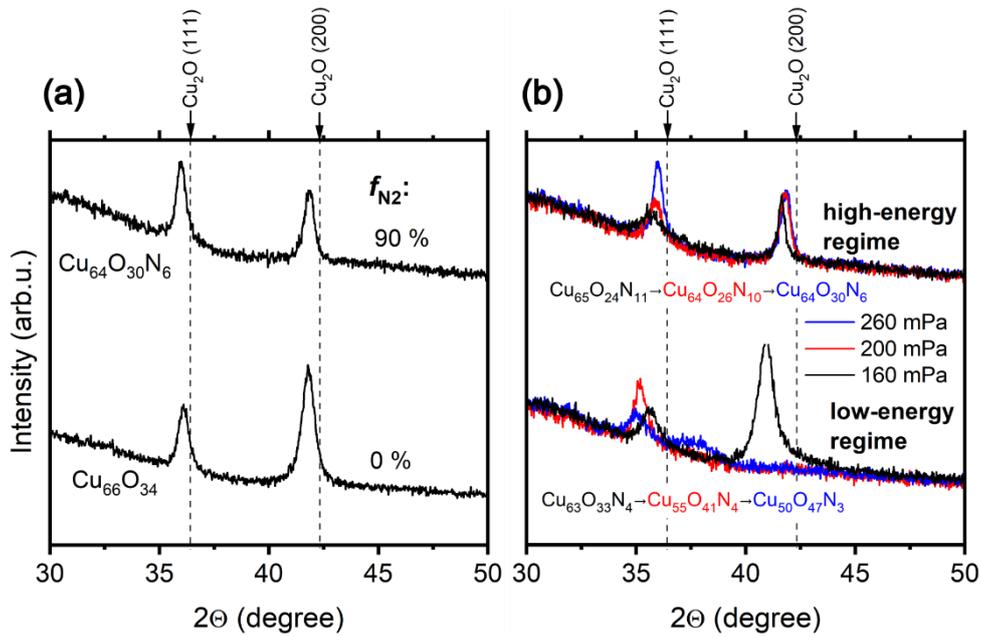

Fig. 4 (a) XRD patterns of the Cu-O-(N) films fabricated in high-energy regime at constant $p_{ox}$ = 260 mPa and different $f_{N2}$ (b) XRD patterns of selected films prepared in low-energy regime (bottom) and high-energy regime (top) at constant $f_{N2}$= 90% and for different values of $p_{ox}$. The position of $Cu_2O$ peaks (stress-free standard) and elemental composition of individual films are also marked for better clarity.

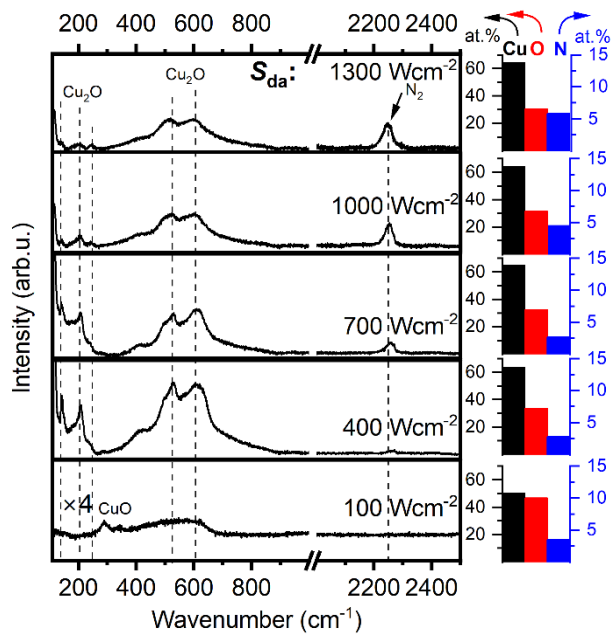

Fig. 5 A Raman spectrum of the Cu-O-N films fabricated at constant $p_{ox}$ =260 mPa and $f_{N2}$ =90% as a function of $S_{da}$. The elemental composition of individual films is plotted for better clarity.

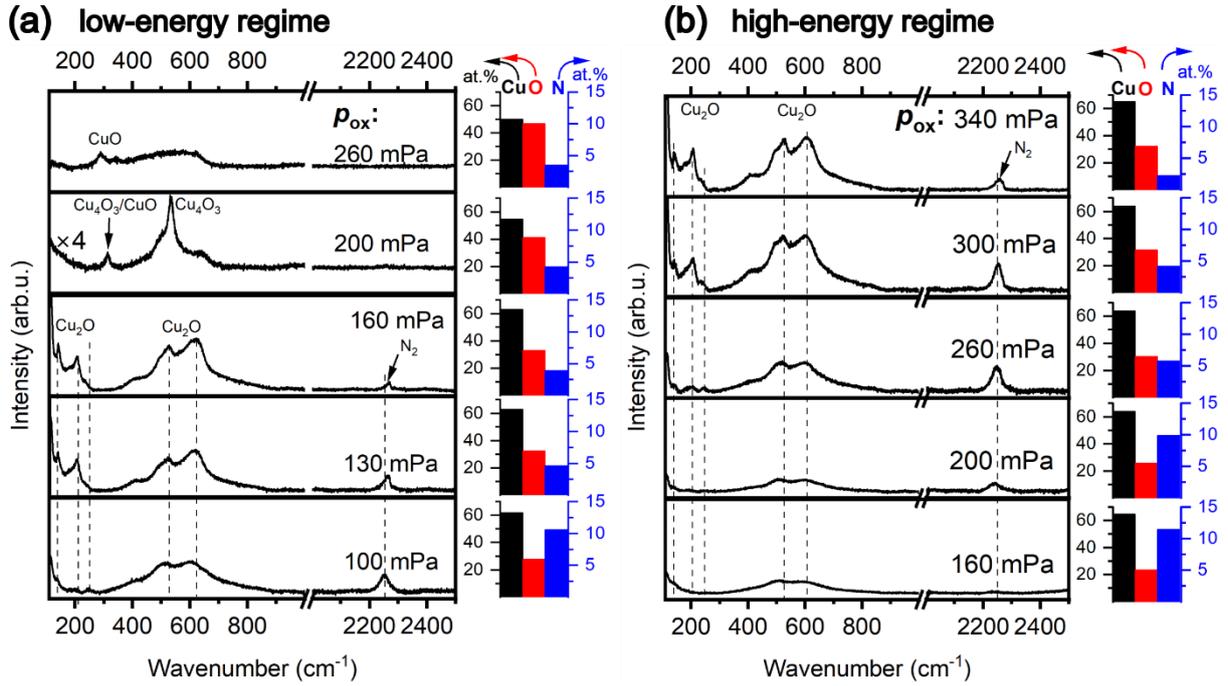

Fig. 6 (a) A Raman spectrum of the Cu-O-N films fabricated in the low-energy regime at constant $f_{N2}$=90% as a function of $p_{ox}$, (b) A Raman spectrum of the Cu-O-N films fabricated in the high-energy regime at constant $f_{N2}$=90% as a function of $p_{ox}$. The elemental composition of individual films is plotted for better clarity.

3.3. Electrical and optical properties

The electrical resistivity and optical band gap determined from Tauc's plots as a function of $f_{N2}$ in the high-energy regime (at $p_{ox}$= 260 mPa) are plotted in **Fig. 7a**. Both the resistivity and the optical band gap decrease with $f_{N2}$. In the case of electrical resistivity, we observe a very significant decrease from the value of $3 \times 10^3$ Ωcm for the nitrogen-free layer to about $5 \times 10^{-2}$ Ωcm for the highest value of $f_{N2}$ = 90%. This value of electrical resistivity surpasses, to our knowledge, all previously published values of electrical resistivity of nitrogen-doped $Cu_2O$ thin films prepared on amorphous substrates. Probably similar resistivity was achieved only in the work of Benz et. Al [16] (although the exact value is not given numerically in the article), however, it was a relatively slow RF deposition on a highly crystalline sapphire substrate. There is an observed decrease in the optical band gap with increasing $f_{N2}$ from approximately 2.50 eV for the nitrogen-free layer to 2.30 eV for the $Cu_{64}O_{30}N_6$ layer prepared at $f_{N2}$=90%.

The gradual decrease in electrical resistivity and optical band gap is also observed if the films prepared at increasing $S_{da}$ (at constant $p_{ox}$ and $f_{N2}$ of 260 mPa and 90%, respectively), see **Fig. 7b**. More precisely, there is an increase in $E_g$ from 2.20 eV to 2.50 eV when $S_{da}$ is increased from 100 Wcm$^{-2}$ to 400 Wcm$^{-2}$. After that, $E_g$ decreases monotonically with $S_{da}$. The initial increase is due to the fact that the layer prepared for the lowest $S_{da}$ = 100 Wcm-$^2$ exhibits CuO structure (which has lower $E_g$ than $Cu_2O$).

The benefit of the high-energy regime for tuning of optoelectric properties is clearly seen in **Fig. 8**. In the low-energy regime, the resistivity is very sensitive to the value of $p_{ox}$ and, mainly due to various phase transitions, changes between 100 mPa and 260 mPa by three orders of magnitude. In the high-energy regime, the resistivity is almost constant and at the same time very low (from ≈ 5 × 10$^{-2}$ Ωcm to ≈ 2 × 10$^{-1}$ Ωcm) in a wide range of conditions from $p_{ox}$ = 160 to 300 mPa. At the same time, the optical band gap changes very significantly, from 2.0 eV for 160 mPa to 2.5 eV for 300 mPa. This independent tuning of the optical band gap can be advantageous, for example, for photovoltaic absorber applications, where it is necessary to precisely tune the band gap edge while maintaining a low electrical resistivity. Another interesting observation is that the $Cu_2O$:N film prepared under the high-energy regime ($p_{ox}$ = 260 mPa, see **Fig. 6b**) containing 5.8 at.% nitrogen exhibits very similar optoelectrical properties ($\rho$ ≈ 5 × 10$^{-2}$ Ωcm, $E_g$ ≈ 2.35 eV) as the film prepared under low-energy regime containing 10.6 at.% of nitrogen ($p_{ox}$ = 100 mPa, see **Fig. 6b**). Although the nitrogen concentration is roughly half for the "high-energy" layer, the intensity of the Raman peak corresponding to the $N_2$ molecule is higher compared with the low-energy layer. This indicates that the fraction of nitrogen in molecular form is higher in the case of "high-energy" film and, in our opinion, still keeps the nitrogen molecule in play as a serious candidate responsible for increasing the hole concentration, see also later in the DFT calculations paragraph. However, further research is still needed.

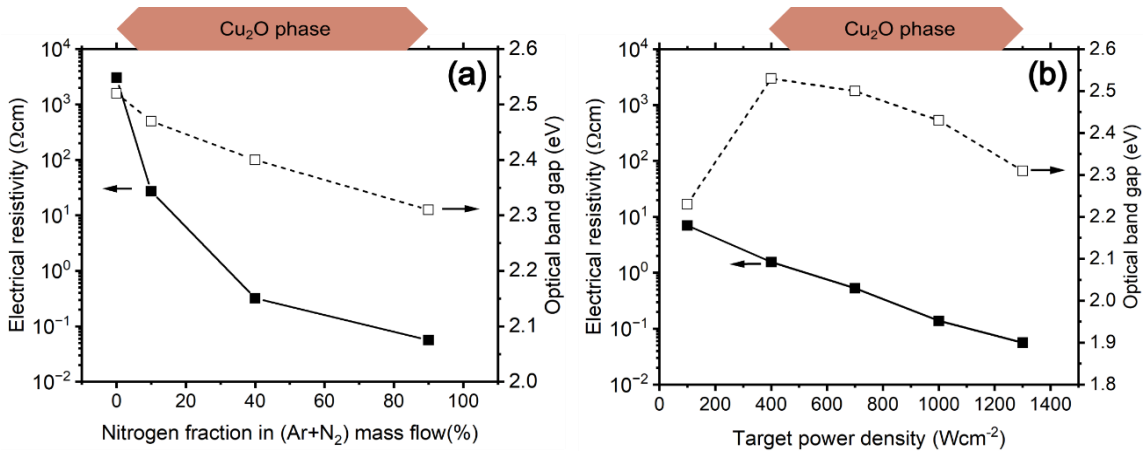

Fig. 7 (a) Electrical resistivity and optical band gap of the Cu-O-N films prepared in high-energy regime at constant $p_{ox}$ = 260 mPa as a function of $f_{N2}$. (a) Electrical resistivity and optical band gap of the Cu-O-N films prepared at constant $p_{ox}$=260 mPa and $f_{N2}$= 90% as a function of $S_{da}$. The intervals when only the Cu$_2$O phase is detected are marked.

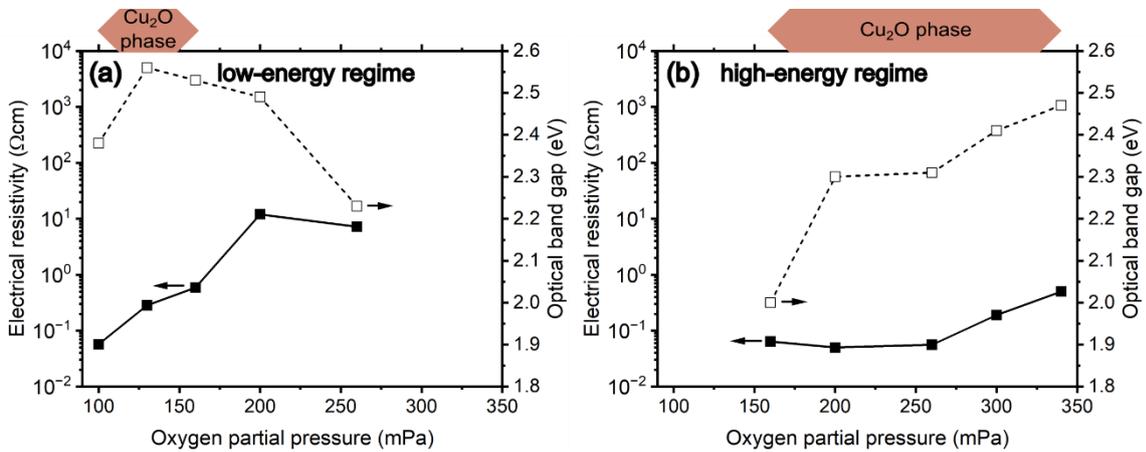

Fig. 8 (a) Electrical resistivity and optical band gap of the Cu-O-N films prepared at constant $f_{N2}$ = 90% as a function of $p_{ox}$ in the low-energy regime, and (b) in the high-energy regime. The intervals when only the Cu$_2$O phase is detected are marked.

*Spectroscopic ellipsometry*

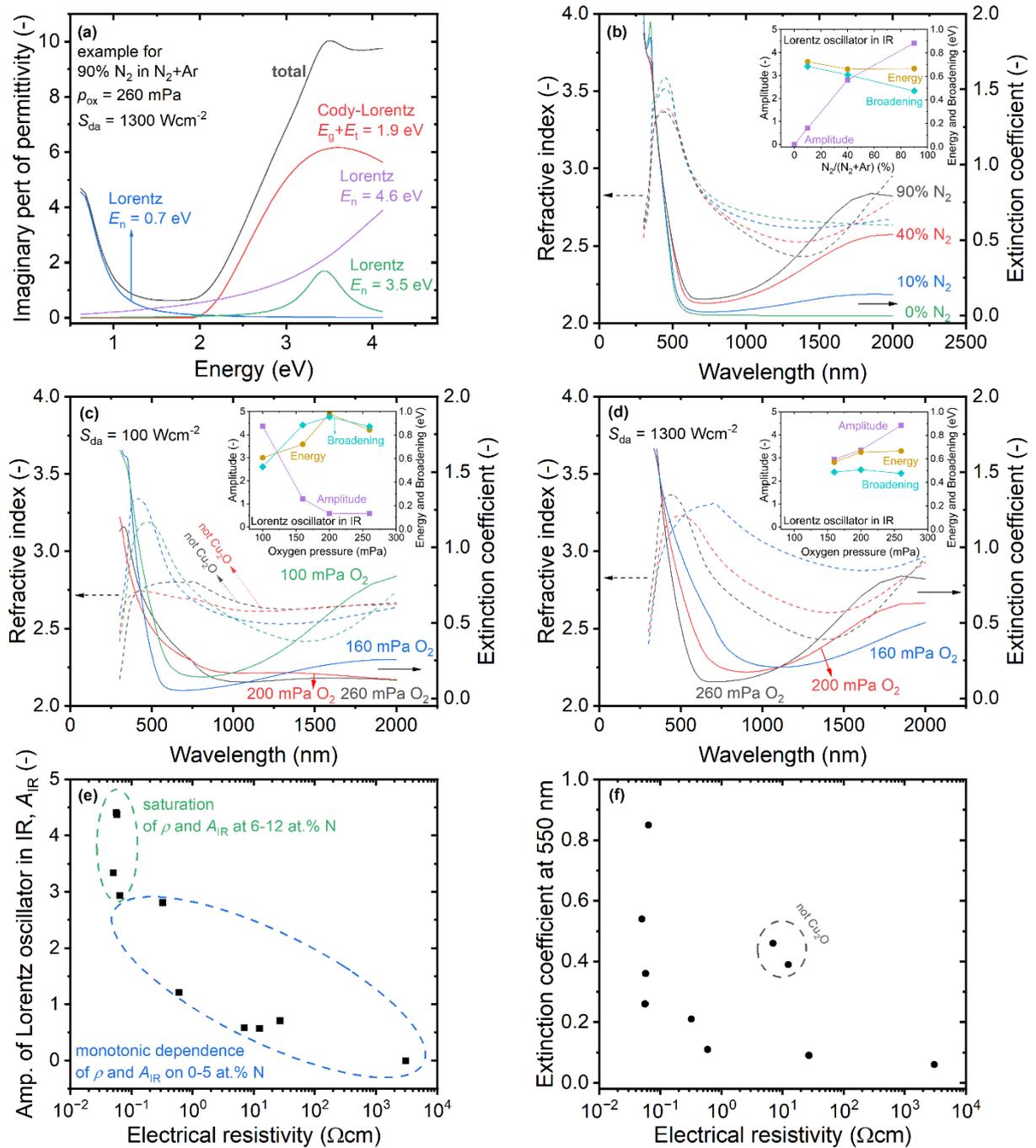

Fig. 9 Panel (a) shows an example of the optical model of Cu-O-N: imaginary part of relative permittivity as a sum of four oscillators discussed in the text. Panels b-d show spectral refractive index (left axes, dashed lines) and extinction coefficient (right axes, solid lines) at $p_{ox}$ = 260 mPa, $S_{da}$ = 1300 Wcm$^{-2}$ and varied $f_{N2}$ (b), $f_{N2}$ = 90%, low-energy regime and varied $p_{ox}$ (c) and $f_{N2}$ = 90%, high-energy regime and varied $p_{ox}$ (d). The insets show parameters of the Lorentz oscillator centred in the IR. Panel e shows the relationship of electrical resistivity and amplitude of the oscillator centred in IR (AIR) for N-poor samples (where both quantities monotonically depend on [N]) as well as N-rich samples (where they almost saturate). Panel f shows the relationship of electrical resistivity and extinction coefficient at $\lambda$ = 550 nm.

Optical properties of selected Cu-O-N materials studied are shown in **Fig. 9**. As illustrated in **Fig. 9a** for one of the samples, the materials have been represented by a sum of (i) the Cody-Lorentz dispersion formula (Eq. 1), having the optical gap $E_g+E_t$ (2.2±0.4 eV for samples dominated by $Cu_2O$) as one of its parameters, with (ii) Lorentz oscillator (Eq. 2) in the infrared, centered at $E_n \approx 0.8\pm0.2$ eV (all samples prepared at a presence of $N_2$, not necessary for pure $Cu_2O$), (iii) Lorentz oscillator in the ultraviolet, centered at $E_n \approx 3.7\pm0.2$ eV (all samples) and (iv) another Lorentz oscillator in the ultraviolet, centered at $E_n \approx 4.7\pm0.5$ eV (all samples dominated by $Cu_2O$, not necessary after the transition from $Cu_2O$ to $Cu_4O_3$ and $CuO$).

Note that although the aforementioned Tauc gaps obtained by spectrophotometry (2.0 - 2.5 eV) are not equal to the $E_g+E_t$ values (1.8-2.6 eV) of $Cu_2O$ obtained by ellipsometry, there is an almost monotonic dependence between these quantities (not shown). The only significant outliers are O-rich samples containing $Cu_4O_3$ and $CuO$ rather than $Cu_2O$, characterized by the least sharp band gap edges and significantly different dispersion of optical constants (high $p_{ox}$ = 260 mPa in the low-energy regime; see two of the dependencies in **Fig. 9c**).

The characteristics of the Lorentz oscillator centred in the infrared are arguably most revealing. The inset in **Fig. 9b** shows that increasing $N_2$ partial pressure and, in turn, increasing N incorporation into the films (**Fig. 1a**) leads to a steeply increasing amplitude of this oscillator ($A_{IR}$) from 0 to ≈ 4.4, at approximately constant position and broadening. Thus, a strong case can be made that this oscillator represents the electronic states inside the band which are localized on nitrogen. The position of this oscillator, $E_n \approx 0.8\pm0.2$ eV, is a measure (upper bound of) the distance of these states from the band gap edge. Similar evolution has also been observed for the oscillator centered in the ultraviolet region at $E_n \approx 4.70\pm5$ eV (not shown).

Increasing the N incorporation indirectly, by decreasing the partial pressure of oxygen, has an effect on the electronic structure of Cu-O-N as well. In the low-energy regime (**Fig. 9c**), at mostly low [N] values (only one sample with >5 at.% N; **Fig. 2a**), the phenomenon is similar to that observed in **Fig. 9b**: $A_{IR}$ increases with decreasing $p_{ox}$ (increasing [N]), once again up to ≈ 4.4. However, in the high-energy regime (**Fig. 9d**), leading to only high [N] values (all samples with >5 at.% N; **Fig. 2b**), $A_{IR}$ does not increase with [N] and even slightly decreases. This indicates that while the incorporation of up to ≈ 5 at.% N preserves the main features of the electronic structure of $Cu_2O$ except the addition of new electronic states inside the band gap, further N incorporation (here up to 12 at.% N) leads to further qualitative changes.

The presented phenomena are examined in **Fig. 9e**, which shows a very good correlation (even including the two samples containing $Cu_4O_3$ and CuO) between two quantities which are both related to the concentration of free charge carriers: $\rho$ and concentration of electronic states inside the band gap in terms of $A_{IR}$. While 0-5% at.% N leads to a practically monotonic dependence of both these quantities on [N], at higher N contents both these quantities saturate close to $\rho \approx 5 \times 10^{-2}$ $\Omega$cm (see also the saturation in **Fig. 8b**) and $A_{IR} \approx 4.4$.

In addition to the spectral dependencies of $n(\lambda)$ and $k(\lambda)$ shown in **Fig. 9b-d**, let us point out a representative and frequently used quantity: $k$ at $\lambda = 550$ nm, $k_{550}$. The relationship between key optoelectronic properties, $\rho$ and $k_{550}$, regardless the N content, is shown in **Fig. 9f**. This time the samples containing $Cu_4O_3$ and CuO constitute outliers. However, the $Cu_2O$-like samples yield a smooth dependence: $k_{550}$ increases with decreasing $\rho$ from 0.06 to 0.26 until the latter reaches the aforementioned saturation value of $5 \times 10^{-2}$ $\Omega$cm. Then, $k_{550}$ can still be varied in a wide range from 0.26 to 0.85 (which of these values is better depends on the application) at approximately constant $\rho$.

*Charge transport properties*

Because it is very complicated or even impossible to measure charge transport properties (i.e. Hall mobility and concentration of major charge carriers) in $Cu_2O$ films heavily doped by nitrogen, we have performed illuminated open-circuit potential (OCP) measurements of selected films. Illuminated Open Circuit Potential (OCP) is the most straightforward photoelectrochemical method for determining the conductivity type of a semiconductor. Illumination above the bandgap generates electron–hole pairs, which are separated by the internal electric field within the depletion region. The minority charge carriers migrate toward the semiconductor–electrolyte interface to facilitate chemical reactions, while the majority charge carriers generate an opposing electric field. Consequently, if the OCP shifts positively (toward more anodic potentials) upon above-bandgap illumination, the material is identified as p-type [31].

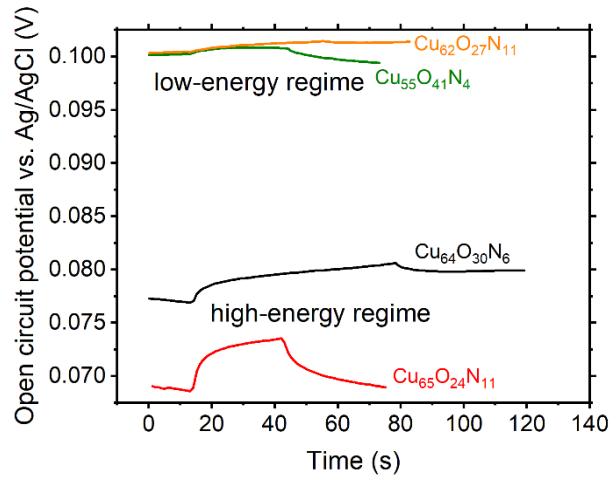

Fig. 10 The open circuit potential for the selected films as a function of time. The elemental composition of individual films are marked for the better clarity.

We measured four films, all prepared with the highest nitrogen content ($f_{N2}$ = 90%). The first two were synthesised under a low-energy regime with oxygen partial pressures of $p_{ox}$ = 100 and 200 mPa, while the second pair was prepared under a high-energy regime ($S_{da}$ = 1300 Wcm$^{-2}$) with oxygen partial pressures of $p_{ox}$ = 160 and 260 mPa. These films were chosen to cover a broad range of conditions and corresponding material properties. All films exhibited moderate photoactivity. Although the OCP shifts were only on the order of millivolts, the consistently positive shifts indicate that all measured films possessed p-type conductivity.

Despite the low-energy regime film at $p_{ox}$ = 100 mPa differing significantly from the other three, including approx. 200 times higher electrical resistivity, distinct elemental and crystalline phase composition, the most pronounced differences in both dark OCP values and illuminated OCP responses were attributable to the regime of deposition (low- or high-energy). Films prepared under high-energy conditions exhibited lower dark OCP values, indicating a greater susceptibility to electron loss and corrosion due to lower electronegativity and/or a reduced ability to form protective oxide layers (passivation). More importantly, both high-energy regime films also demonstrated stronger and faster illuminated OCP responses. This indicates a more pronounced p-type semiconductor character, where hole conduction dominates. Such behaviour suggests reduced charge carrier recombination due to a lower density of defect states (recombination centres), thereby enabling higher charge carrier mobility.

### 3.4. DFT calculations

**Table 1** confirms that all nitrogen-doped configurations in $Cu_2O$ have positive formation energies, indicating endothermic incorporation and underlining the advantage of the non-equilibrium technique used for the preparation of these configurations. Among them, the most favorable configurations correspond to nitrogen atoms substituting oxygen atoms and nitrogen molecules ($N_2$) substituting copper atoms. These were selected for further analysis. The calculated band gap of pristine $Cu_2O$ is 0.51 eV (**Fig. 11a**), consistent with previous DFT-GGA reports (0.55–0.70 eV, A. Larabi [32]), though significantly lower than the experimental value of 2.1-2.6 eV [2,3] due to the known band gap underestimation (albeit at preserved trends) by GGA functionals. Upon nitrogen doping, the band gap increases to 0.89 eV for $N_2$ doping at Cu sites and 0.72 eV for N doping at O sites (**Fig. 11b** and **Fig. 11c**, respectively). This band gap widening is accompanied by the key result of the DFT calculations: shift of the Fermi level toward the valence band in both cases, indicating increased hole carrier concentration (see the lowered electrical resistivity in **Fig. 7** and **Fig. 8**) and further supporting the desired p-type character of the material. Additionally, although the new electronic states arising from nitrogen doping (projected density of states in **Fig. 11e** and **Fig. 11f**) are not exclusively localized near the Fermi level, **Fig. 11f** shows disproportionately strong role of N at the Fermi level and just below it at the top of the valence band (but not at the bottom of the conduction band). This is consistent with the spectroscopic ellipsometry. While the results of ellipsometry (**Fig. 9**) could be in part explained by new electronic states close to the edge of some band after N incorporation (a new Lorentz oscillator centred in the infrared), DFT allows us to specify that they are close to the edge of the valence band.

Table 1 Formation energies of N atom and $N_2$ molecule at various sites in the $Cu_2O$ lattice.

| Sites | $N$-$Cu_2O$ (O-site) | $N$-$Cu_2O$ (Cu-site) | Interstitial position (N atom) | $N_2$-$Cu_2O$ (O-site) | $N_2$-$Cu_2O$ (Cu-site) | Interstitial position ($N_2$) |
|---|---|---|---|---|---|---|
| $E_f$(eV) | 2.26 | 4.87 | 3.00 | 4.58 | 1.67 | 2.86 |

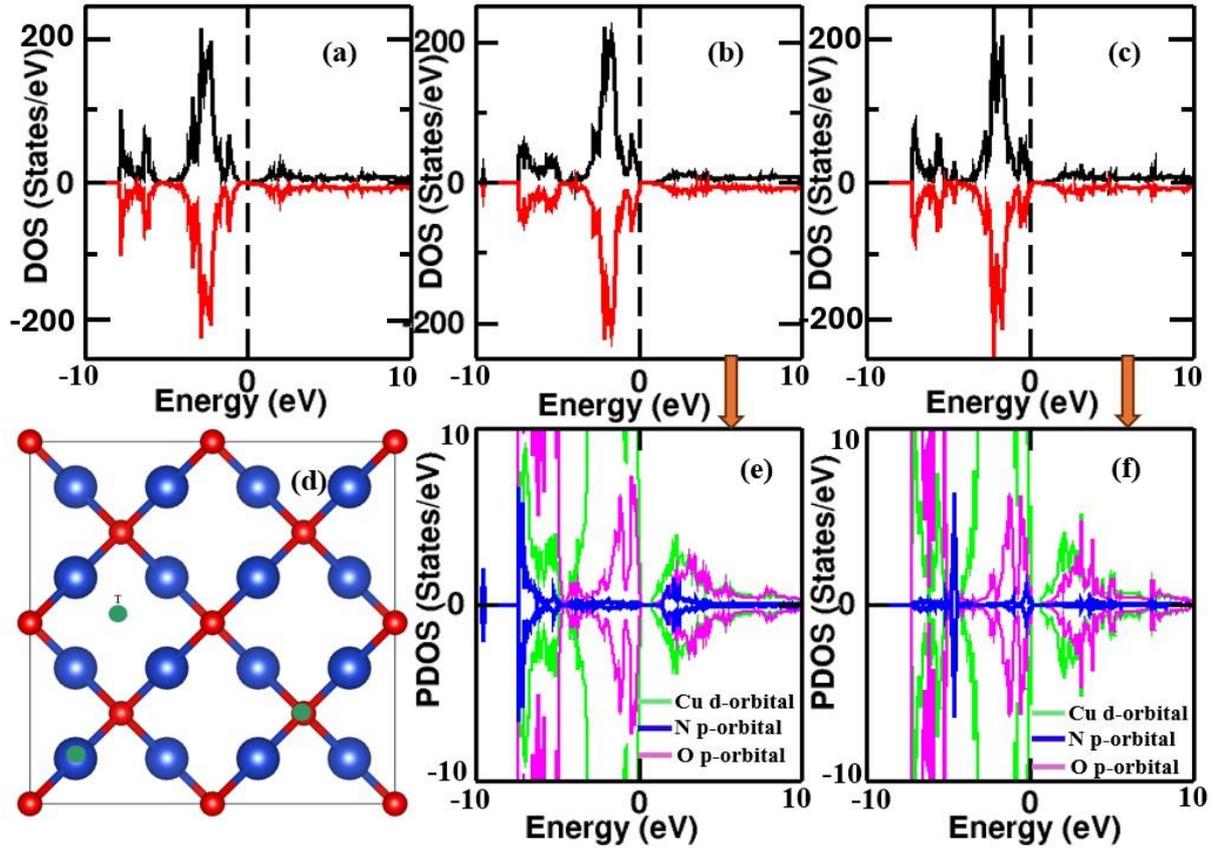

Fig. 11 (a) Total DOS of pure $Cu_2O$, (b) $N_2$-doped $Cu_2O$ (Cu substitution, the most energetically preferred $N_2$ site), (c) N-doped $Cu_2O$ (O substitution, the most energetically preferred N site), (d) optimized $2 \times 2 \times 2$ $Cu_2O$ supercell structure (Cu: blue, O: red, N: green), (e) projected density of states of $N_2$-doped $Cu_2O$ (Cu substitution), and (f) projected density of states of N-doped $Cu_2O$ (O substitution). The Fermi level is set to 0 eV.

4. Conclusions

The nitrogen-doped p-type cuprous oxide thin films with very low electrical resistivity down to $\approx 5 \times 10^{-2}$ $\Omega$cm were successfully fabricated. The main conclusion of this work could be summarized as follows:

- High-energy regime (i.e. discharge at high pulse-averaged target power density of 1300 Wcm$^{-2}$) is more suitable for the preparation of $Cu_2O$-like films and enables continuous replacement of oxygen by nitrogen ($Cu_{65}O_{32}N_4 - Cu_{65}O_{23}N_{12}$) just by changing the oxygen partial pressure during the deposition without undesirable transition to $Cu_4O_3$ or CuO. This results in the independent tuning of optical properties

(for instance, $E_g$ from 2.0 to 2.5 eV) while maintaining a very low electrical resistivity of $5 \times 10^{-2}$ Ωcm - $2 \times 10^{-1}$ Ωcm.

- The results from spectroscopic ellipsometry, supported by DFT calculations, clearly show that nitrogen is responsible for the new acceptor states in the form of substitution of atomic N for O or substitution of $N_2$ molecule for Cu

- The nitrogen molecule seems to be more important in the films grown under a high-energy regime and still is at least partially responsible for the increase in hole concentration in $Cu_2O:N$ films

**CRediT authorship contribution statement**

**Jiří Rezek**: Conceptualization, Methodology, Investigation, Visualization, Writing – original draft, Writing – review & editing. **Jan Koloros:** Investigation, Experiment **Jiří Houška:** Methodology, Investigation, Visualization, Writing – original draft. **Radomír Čerstvý:** Experiment. **Stanislav Haviar** Investigation. **David Kolenatý**: Experiment, Writing – original draft. **Jemal Yimer Damte:** Investigation, Visualization, Writing – original draft. **Pavel Baroch:** Supervising, Conceptualization.

**Acknowledgement**

This work was supported by the project Quantum materials for applications in sustainable technologies (QM4ST), funded as project No. CZ.02.01.01/00/22_008/0004572 by Programme Johannes Amos Comenius, call Excellent Research.